\documentclass{layout/tudelft-aiaa}

\title{Lossless photon flux sensing methodology for \\ EUV and Soft X-ray metrology systems}
\author{S. Weerdenburg}
\affil{Optics Research Cluster, ImPhys Department, Delft University of Technology, The Netherlands}
\author{R. Horsten}
\affil{Optics Research Cluster, ImPhys Department, Delft University of Technology, The Netherlands}
\author{W.M.J.M. Coene}
\affil{Optics Research Cluster, ImPhys Department, Delft University of Technology, The Netherlands}
\affil{ASML Research, Veldhoven, The Netherlands}

\begin{document}

\AlwaysPagewidth{

\maketitle


\begin{abstract}
We demonstrate a simple, lossless method for monitoring photon flux in short-wavelength metrology systems, with a particular focus on applications using high harmonic generation (HHG) light sources. In HHG-based metrology, where photon scarcity often limits precision and efficiency, the ability to monitor flux without sacrificing photons is critical. This demonstrated approach measures a photon-induced current in a metallic EUV filter, generated by primary and secondary electrons released via the photoelectric effect during exposure to high-energy photons. By integrating this current over a set exposure period, we obtain a build-up charge directly correlated with the incident photon flux. This measurement provides a reliable, real-time indicator of photon flux, allowing for precise normalization in metrology experiments without introducing additional optical losses. The method leverages any conductive component that are typically present in extreme ultraviolet (EUV) metrology setups, making it a cost-effective and straightforward addition to existing systems.
\end{abstract}

}

\section{Introduction}
Many optical metrology systems can be generalized into three basic modules: illumination, sample interaction, and detection. These types of metrology systems rely on the photon flux stability of the light source as it is not possible to differentiate flux change from a change in the sample by just using the diffracted or scattered light. Therefore, the photon flux illuminating the sample has to be monitored to normalize the measured intensities in the detector after interaction with the sample. 

In most cases, this can be solved by picking off a small portion of the beam utilizing a beamsplitter or secondary reflections. This approach is quite viable and sufficient for situations where photons are abundant or where beamsplitters are relatively efficient. Unfortunately, the latter two requirements are often not met in short-wavelength systems such as table-top Extreme UV (EUV) light sources based on High Harmonic Generation (HHG). Even though these HHG systems have become more efficient over time \cite{Popmintchev2010,Rothhardt_2018}, the generated short-wavelength photons remain rather scarce and thus quite precious. Additionally, beam splitters in this wavelength regime are relatively inefficient. \cite{Maki:10} 

This tutorial demonstrates a cheap and straightforward method based on ejected primary and secondary electrons (from the
photo-electric effect) to sense the photon flux with components inherently found in most EUV metrology systems based on high harmonic generation. Since this technique utilizes components that are already present within such a beamline, it does not result in any losses in the photon flux as shown in mirrors before. \cite{grilj2013beamline} 

\section{Methodology}
\subsection{Extreme UV Metrology Beamline}
Without loss of generality, we consider here an EUV beamline that is configured for lensless imaging in reflection mode at 18 nanometer wavelength based on high harmonic generation. The compact HHG system (from Active Fiber Systems \cite{Tschernajew:20}) is pumped by a 28 femtosecond, 110 µJ infrared ($\lambda$ = 1030 nm) Ytterbium fiber laser with a repetition rate of 600 kHz, resulting in an average power of 68 W after pulse compression. Higher harmonics are produced by focusing the drive laser into a 10 bar pressurized gas jet of Argon. These higher harmonics are generated down to the Soft X-Ray regime, with a single-harmonic photon flux of 2 $\cdot$ $10^{11}$ photons per second within 1 percent of bandwidth at the wavelength of our interest $\lambda$ = 18 nm or 68 eV. This specific beamline has been configured for coherent diffractive imaging, or lensless imaging. \cite{shao2024wavelength,weerdenburg2023euv} 

\begin{figure}[htbp]
\centering\includegraphics[width=0.85\textwidth]{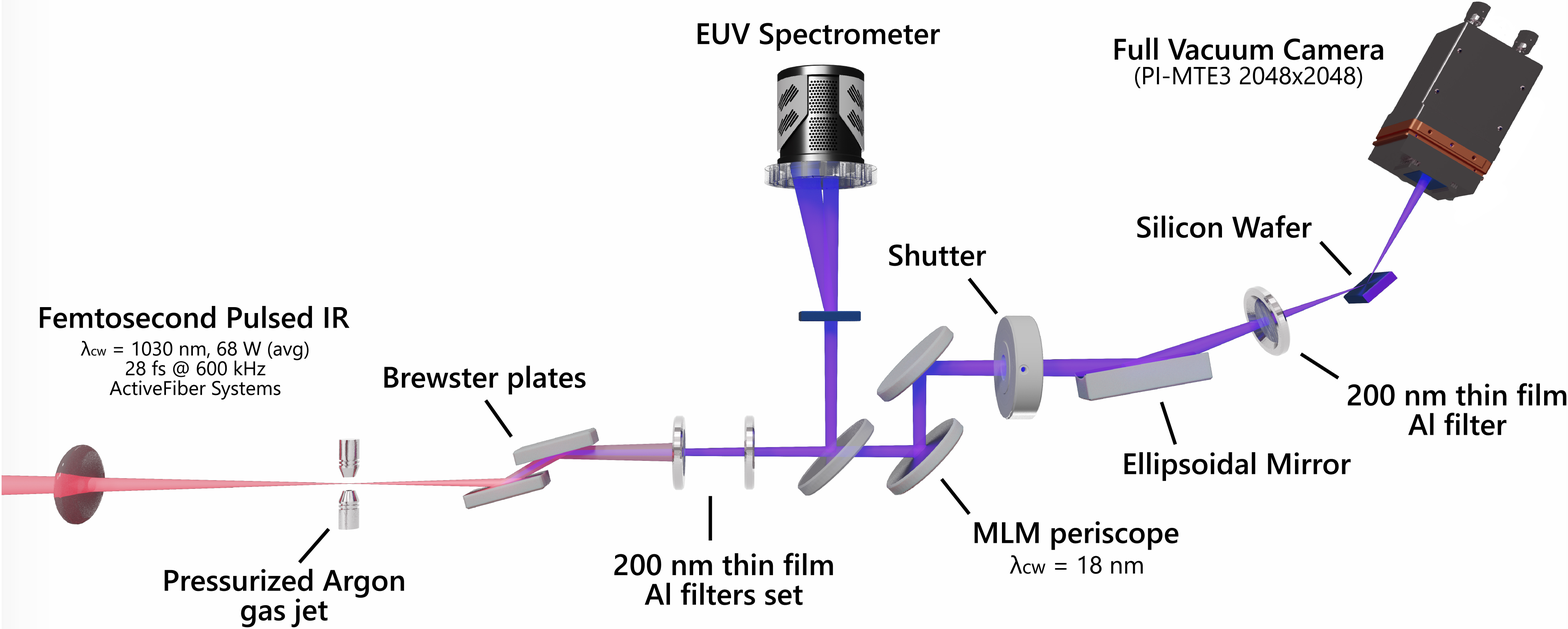}
\caption{The beamline is configured for lensless microscopy in the Extreme UV/Soft X-Ray regime in reflection mode. Downstream, past the HHG source, we select a single harmonic through multi-layer mirrors with a center wavelength of 18 nm. The last metallic (200 nm Al) filter, located between the ellipsoidal mirror and a silicon wafer, is used to measure a photon-induced current.}
\label{fig:setup}
\end{figure}

\begin{figure*}[!htbp]
\centering\includegraphics[width=\textwidth]{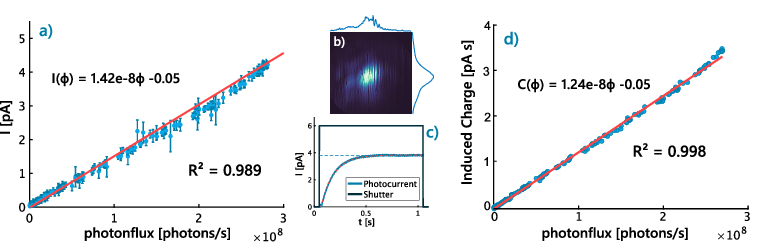}
\caption{A "steady-state current" is found by fitting Eq.(\ref{eq:risetime}) into each signal fragment during a complete exposure as indicated in \textbf{c)} and compared to the photon flux of the reflected beam measured on the CCD in \textbf{b)}. Repeating this for a different photon flux, up to 3 $\cdot 10^{8}$ photons per second, results in the linear relation between the photon current and photon flux as shown in \textbf{a)}. However, the flux varies over time, yielding a significant steady-state current error, hence the larger error bars. Integrating the current during the entire exposure, including the rise time of the detector, is able to compensate these effects \textbf{d)}.}
\label{fig:SingleHarmonic_Filter}
\end{figure*}

Given the typical low conversion efficiency of the HHG process, the IR beam has to be well separated from the higher harmonics to be utilized in a metrology experiment. Our beamline does this in three stages; A dichroic mirror with an aperture of 2 mm in it's center (located in the first module) is utilized to clip the 4.7 mm wide IR beam. The EUV beam is still able to pass, due to the smaller beam divergence compared to the IR beam.

The second module consists of two rejection mirrors set to the Brewster angle for the infrared beam while still reflecting higher harmonics with a combined efficiency of 54\%. The residual infrared light is blocked in the last module, by a set of 200 nm thick aluminum freestanding foils. Aluminium has a transmission cut-off at 73 eV ($\lambda_{\text{cutoff}}$ = 17 nm) acting as a long-pass filter for our higher harmonics. 

A single harmonic is selected by a set of multi-layer mirrors (MLM), which reflect 18 nm with a combined peak efficiency of 21\% and FWHM of 0.6 nm. Further downstream of the beamline, a Ruthenium-coated grazing incidence ellipsoidal mirror focuses the beam on a sample of interest for metrology purposes. In this case, the sample would be a silicon wafer with 2D or 3D patterned structures. The reflected light from the silicon wafer is captured by a 2048 by 2048 pixel full-vacuum camera (PI-MTE3) with a pixel size of 15 micrometer.

\subsection{Photon flux sensitive optical elements}

Our SXR metrology beamline contains application-specific optical elements. However, generic components can be found in similar beamlines such as the metallic transmission filters, and grazing incidence mirrors \cite{Ku:16,Porter:17,WACHULAK2012102,Tadesse:16,Jansen:16}. All these optical components enable us to measure the photon flux impinging on their conductive surfaces. The high-energy photons are energetic enough to release electrons from the surface through primary and secondary electrons. These electrons will end up somewhere within the vacuum chamber and the majority will end up in a common ground. Therefore, a photon-induced current can be measured on the conductive surface of the mirrors and filters which can be used as a normalization reference in metrology experiments.

As this approach does not yield any spectral information, just like any photodiode, the location of the sensing optical object would ideally as close as possible to the application. The optimal location in our beamline would be the last optical element, i.e. the ellipsoidal mirror, as only a single harmonic hits this surface. All the light reflected from the mirror propagates towards the sample of interest. However, in this tutorial, we demonstrate this method on a 200 nm thin aluminum transmission filter instead as it is rather easy to integrate and is a more generic use case. Our beamline already has a set of two metallic filters, which are needed to suppress the IR light sufficiently but for this demonstration, we add a third aluminum filter in between the mechanical shutter and the sample as shown in Figure \ref{fig:setup}. In this way, we illuminate the filter with a single harmonic rather than with a broadband beam. 

The aluminum filter is only 200 nm thick and thus delicate, making it hard to attach an electrical contact directly onto the foil. Fortunately, the foil is mounted on a steel ring which is sandwiched between two steel washers, on which we directly soldered an electrical contact with a single core wire (i.e. floating ground). The wire is attached to vacuum compatible coax feed through. The small current from the foil needs to be amplified in order to be measured for which we use a variable low-noise preamplifier (SRS Stanford Research Systems SR570), which output is sampled at 30 kHz by an Analog-to-Digital converter (ADC, NI 9234). Depending on the expected photonflux, one would choose their preamplifier. In our case we expected to have currents in the picoampere range, putting steep requirements on the preamplifier. 

\section{Experimental Evaluation}

The metallic filter is exposed to a single harmonic with a center wavelength of 18 nanometer as soon as the mechanical shutter opens. It takes some time for the current in the foil to settle to a "steady-state current" $I_{\text{max}}$ of 4.5 pA, see Fig. \ref{fig:SingleHarmonic_Filter}c, with an average time constant $\tau$ of 144 $\pm$ 11 ms. The time constant and steady-state current are defined according to Eq. \ref{eq:risetime}. In order to calibrate the measured current to a photon flux, we use the reflection of a bare silicon sample, which is captured by the CCD placed at the end of the beamline, see Figure \ref{fig:SingleHarmonic_Filter}b. After the exposure of one second, the integrated counts can be converted to a photon flux. This value is used as a reference to the measured current and is found to be $2.69 \cdot 10^8$ photons per second for 3.9 pA. Note that the captured image of the CCD is after reflection with a silicon sample, meaning that the actual flux hitting the foil is higher than measured with the CCD.

\begin{equation}
    I(t) = I_{\text{max}}\Big(1 - \exp{\Big[-\frac{t-t_0}{\tau}\Big]}\Big)
    \label{eq:risetime}
\end{equation}

By gradually physically obscuring the EUV beam, we can vary the photon flux impinging on the filter. However, as we induce a significant change in the flux, we cannot find a 'steady state' current accurately by fitting Eq. \ref{eq:risetime} per exposure.

The relatively larger error bars on specific data points in Fig. \ref{fig:SingleHarmonic_Filter}a, are indication of such a significant change in the photon flux in that exposure. This can be simply solved by integrating the photon-induced current during the entire exposure, meaning we find a total build-up charge on the foil. Furthermore, a linear relation is found between the build-up charge and the photon flux, see Fig. \ref{fig:SingleHarmonic_Filter}d, with a coefficient of determination $R^2$ of 0.998.

Using an bare wire without shielding to connect the preamplifier and the sensor while measuring these small currents, leads to parasitic signals from external sources such as the vibrations induced by the molecular turbo-pumps or the cooling system. These vibrations occur at significantly higher frequencies than typical exposure times for short-wavelength metrology experiments. Thus a simple low-pass filter can remove these types of parasitic signals. 

\begin{figure}[htbp]
\centering\includegraphics[width=0.4\textwidth]{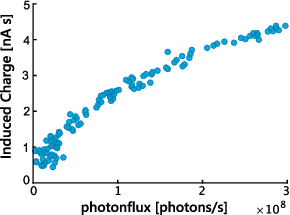}
\caption{A non-linear response is observed when reducing the gas flow in the gas jet when the first filter, before harmonic filtering, is measured. The contribution of other harmonics in the photon current start to dominate relative to the harmonic of interest.}
\label{fig:nonlinearresponse}
\end{figure}

Additionally, the mechanical shutter is driven by an electromagnetic coil to actuate the shutter blades. This coil induces a significant parasitic signal in the unprotected wire due to electromagnetic coupling. The inductive parasitic signal contributes 121 $\pm$ 8 ms to the total time constant of the sensor found in Fig. \ref{fig:SingleHarmonic_Filter}c of 140 ms. These external signals can be reduced by replacing the bare wire with a shielded coaxial wire.

This experiment is repeated on the first aluminum filter in the beamline to increase the photon flux (as it is now exposed to all harmonics) and reduce parasitic currents induced by the shutter. The flux is varied by constricting the gas flow towards the gas jet while measuring the photon flux of a single harmonic (i.e. 18 nanometer)in the CCD plane. Being exposed to all the harmonics, increased the photon-induced current up to 1000 times compared to a single harmonic.

The linear behavior, between the photon flux and build-up charge, we observed for a single harmonic is disrupted by the non-linear behavior of the other produced harmonics when reducing the gas flow and the additional infrared light, see Figure \ref{fig:nonlinearresponse}. The non-linearity measured at that part in the beamline would make it less viable for normalization purposes in low-light ($\lesssim10^8$ photons/s) conditions for a single harmonic.

\section{Conclusion}
We demonstrated a methodology to retrieve a measure of the photon flux in a short-wavelength high harmonic generation metrology beamline, without affecting the photon efficiency of the beamline itself. By illuminating optical elements already present in a typical short-wavelength system, we can measure a photon induced current and charge through primary and secondary electrons. A linear relation has been observed between the induced current and flux when exposed to a single harmonic, leading to a coefficient of determination of 0.998. Therefore this approach enables flux normalization of a single harmonic in short-wavelength metrology systems and encourage the implementation of such sensors in metrology systems.


\bibliography{article}

\end{document}